\documentclass[sigconf]{acmart}

\copyrightyear{2021} 
\acmYear{2021} 
\setcopyright{acmcopyright}\acmConference[SIGIR '21]{Proceedings of the 44th International ACM SIGIR Conference on Research and Development in Information Retrieval}{July 11--15, 2021}{Virtual Event, Canada}
\acmBooktitle{Proceedings of the 44th International ACM SIGIR Conference on Research and Development in Information Retrieval (SIGIR '21), July 11--15, 2021, Virtual Event, Canada}
\acmPrice{15.00}
\acmDOI{10.1145/3404835.3463089}
\acmISBN{978-1-4503-8037-9/21/07}

\usepackage{booktabs} 
\usepackage{multirow}
\usepackage{graphicx}
\usepackage{balance}
\usepackage{subfigure}
\usepackage{bm}
\usepackage{amsmath}
\usepackage{amsfonts}
\usepackage{graphicx}

\usepackage[ruled,vlined]{algorithm2e}
\usepackage{algpseudocode}

\newcommand{\squishlist}{
 \begin{list}{$\bullet$}
  { \setlength{\itemsep}{0pt}
     \setlength{\parsep}{3pt}
     \setlength{\topsep}{3pt}
     \setlength{\partopsep}{0pt}
     \setlength{\leftmargin}{1.5em}
     \setlength{\labelwidth}{1em}
     \setlength{\labelsep}{0.5em} } }

\newcommand{\squishlisttwo}{
 \begin{list}{$\bullet$}
  { \setlength{\itemsep}{0pt}
     \setlength{\parsep}{0pt}
    \setlength{\topsep}{0pt}
    \setlength{\partopsep}{0pt}
\setlength{\leftmargin}{2em}
\setlength{\labelwidth}{1.5em}
\setlength{\labelsep}{0.5em} } }

\newcommand{\squishend}{
\end{list}  }

\setcopyright{acmcopyright}

\begin{document}
\title{Sequential Recommendation for Cold-start Users with \\Meta Transitional Learning}

\author{Jianling Wang}
\affiliation{%
  \institution{Texas A\&M University}
}
\email{jlwang@tamu.edu}

\author{Kaize Ding}
\affiliation{%
  \institution{Arizona State University}
}
\email{kaize.ding@asu.edu}

\author{James Caverlee}
\affiliation{%
  \institution{Texas A\&M University}
}
\email{caverlee@tamu.edu}
\fancyhead{}
\begin{abstract}

 A fundamental challenge for sequential recommenders is to capture the sequential patterns of users toward modeling how users transit among items. In many practical scenarios, however, there are a great number of cold-start users with only minimal logged interactions. As a result, existing sequential recommendation models will lose their predictive power due to the difficulties in learning sequential patterns over users with only limited interactions. In this work, we aim to improve sequential recommendation for cold-start users with a novel framework named MetaTL, which learns to model the transition patterns of users through meta-learning. Specifically, the proposed MetaTL: (i) formulates sequential recommendation for cold-start users as a few-shot learning problem; (ii) extracts the dynamic transition patterns among users with a translation-based architecture; and (iii) adopts meta transitional learning to enable fast learning for cold-start users with only limited interactions, leading to accurate inference of sequential interactions. 
 
 
\end{abstract}

%


\ccsdesc[500]{Information systems}
\ccsdesc[500]{Information systems~Recommender systems}

\keywords{Recommendation Systems, Cold-start, Meta-learning}

\maketitle

\section{Introduction}



Recommendation systems play an essential role in connecting users with content they enjoy on many large-scale online platforms. Different from traditional recommendation systems \cite{NGCF19,rendle2009bpr,he2017neural} that model the general preferences of users, sequential recommendation systems \cite{hidasi2015session,kang2018self,tang2018personalized,yuan2019simple,hidasi2018recurrent,wang2020next} aim to infer the dynamic preferences of users from their behavioral sequences.


And yet, in many real-world scenarios, sequential recommenders may face difficulty in dealing with new users who have only limited interactions with the system, leading to inherently long-tailed interaction data~\cite{yin2020learning}. Ideally, an  effective recommendation system should be able to recommend items to new users who have only a few interactions with items. However, most existing sequential recommenders are not designed to handle such cold-start users due to the difficulty of characterizing user preferences with limited data. Since new users may migrate away from the platform if they receive poor recommendations initially, \textit{how to capture the preference of these cold-start users} becomes a critical question for building a satisfactory recommendation service.


Though quite a few cold-start recommendation methods have been proposed, most require side information~\cite{lee2019melu, li2019zero, zhu2020recommendation} or knowledge from other domains~\cite{mirbakhsh2015improving,kang2019semi,bi2020dcdir} during training, and commonly treat the user-item interactions in a static way. In contrast, cold-start sequential recommendation targets a setting where no additional auxiliary knowledge can be accessed due to privacy issues, and more importantly, the user-item interactions are sequentially dependent. A user’s preferences and tastes may change over time and such dynamics are of great significance in sequential recommendation. Hence, it is necessary to develop a new sequential recommendation framework that can distill short-range item transitional dynamics, and make fast adaptation to those cold-start users with limited user-item interactions.


In this work, we propose a new meta-learning framework called \textit{MetaTL} for tackling the problem of cold-start sequential recommendation.
In order to improve the model generalization capability with only a few user-item interactions, we reformulate the task of cold-start sequential recommendation as a few-shot learning problem. Different from existing methods that directly learn on the data-rich users, MetaTL constructs a pool of few-shot user preference transition tasks that mimic the targeted cold-start scenarios, and progressively learns the user preferences in a meta-learning fashion. Moreover, we build the proposed framework on top of a translation-based architecture, which allows the recommendation model to effectively capture the short-range transitional dynamics. This way the meta-learned sequential recommendation model can extract valuable transitional knowledge from those data-rich users and make fast adaptation to cold-start users to provide high-quality recommendations. The contributions of this work can be summarized as follows:
\squishlist
\item We explore the challenging problem of sequential recommendation for cold-start users without relying on auxiliary information and propose to formulate it as a few-shot learning problem.   
\item We develop a novel meta-learning paradigm -- \textit{MetaTL} to model the transition patterns of users, which can make fast adaption for cold-start users in inferring their sequential interactions. 
\item With extensive experiments on three real-world datasets, we verify the effectiveness of the proposed MetaTL in cold-start sequential recommendation and shows that it can bring in 11.7\% and 5.5\% improvement compared with the state-of-the-art in sequential recommendation and cold-start user recommendation. 
\squishend









\section{Related Work}

\smallskip
\noindent\textbf{Sequential Recommendation.} One of the first approaches for sequential recommendation is the use of Markov Chains to model the transitions of users among items \cite{rendle2010factorizing}. More recently,  TransRec \cite{he2017translation} embeds items in a ``transition space'' and learns a translation vector for each user. With the advance in neural networks, many different neural structures including Recurrent Neural Networks \cite{hidasi2018recurrent,hidasi2015session}, Convolutional Neural Networks \cite{tang2018personalized,yuan2019simple}, Transformers \cite{kang2018self,sun2019bert4rec} and Graph Neural Networks \cite{wu2019session,wang2021session}, have been adopted to model the dynamic preferences of users over their behavior sequences. While these methods aim to improve the overall performance via representation learning for sequences, they suffer from weak prediction power for cold-start users with short behavior sequences. 

\smallskip
\noindent\textbf{Meta-learning .} This line of research aims to learn a model which can adapt and generalize to new tasks and new environments with a few training samples. To achieve the goal of ``learning-to-learn'', there are three types of different approaches. Metric-based methods are based on a similar idea to the nearest neighbors algorithm with a well-designed metric or distance function \cite{vinyals2016matching}, prototypical networks \cite{snell2017prototypical,ding2020graph} or Siamese Neural Network \cite{koch2015siamese}. Model-based methods usually perform a rapid parameter update with an internal architecture or are controlled by another meta-learner model \cite{santoro2016meta}. As for the optimization-based approaches, by adjusting the optimization algorithm, the models can be efficiently updated with a few examples \cite{nichol2018first,finn2017model,ding2021few}. In this work, we explore how to design an effective approach to handle the cold-start sequential recommendation for short sequences. 

\smallskip
\noindent\textbf{Cold-start Recommendation.} Under the complete cold-start setting with no historic interaction available for new users or items, previous works usually learn a transformation between the auxiliary information with the well-trained latent factors \cite{van2013deep,gantner2010learning,li2019zero}. Under the incomplete cold-start setting, Dropoutnet utilizes Dropout layer to simulate the data missing problem \cite{volkovs2017dropoutnet}. Meanwhile, meta-learning has been applied to train a model tailored for cold-start cases.  To solve the user cold-start problem, MetaRec \cite{vartak2017meta} proposes a meta-learning strategy to learn user-specific logistic regression. There are also methods including MetaCF \cite{wei2020fast}, Warm-up \cite{pan2019warm} and MeLU \cite{lee2019melu}, adopting Model-Agnostic Meta-Learning (MAML) methods to learn a model to achieve fast adaptation for cold-start users. However, none of them are designed with the dynamics of user preferences in mind (as is the case in sequential recommendation). 


\section{The Proposed Model}
In this section, we introduce the details of the proposed MetaTL. In essence, the design of MetaTL aims to answer the following research questions: \textbf{RQ1}: How to enable the model to transfer knowledge from data-rich users to cold-start users? \textbf{RQ2}: How do we capture the short-range transition dynamics in user-item interaction sequences? and  \textbf{RQ3}: How to optimize the meta-learner for making accurate recommendations for cold-start users?

\subsection{Problem Setup}
Let $\mathcal{I}=\{i_1, i_2, ..., i_P\}$ and $\mathcal{U}=\{u_1, u_2, ..., u_G\}$ represent the item set and user set in the platform respectively. Each item is mapped to a trainable embedding associated with its ID. There is no auxiliary information for users or items. In sequential recommendation, given the sequence of items $Seq_u = (i_{u,1}, i_{u,2},..., i_{u,n})$ that user $u$ has interacted with in chronological order, the model aims to infer the next interesting item $i_{u,{n+1}}$. That is to say, we need to predict the preference score for each candidate item based on $Seq_u$ and thus recommend the top-N items with the highest scores. 

In our task, we train the model on $\mathcal{U}_{train}$, which contains users with various numbers of logged interactions. Then given $u$ in a separate test set $\mathcal{U}_{test}$, $\mathcal{U}_{train} \cap \mathcal{U}_{test} = \emptyset$, the model can quickly learn user transition patterns according to the $K$ initial interactions and thus infer the sequential interactions. Note that the size of a user’s initial interactions (i.e., $K$) is assumed to be a small number (e.g., 2, 3 or 4) considering the cold-start scenario. 

\begin{figure}
\vspace{-0.1in}
\includegraphics[width=0.43\textwidth]{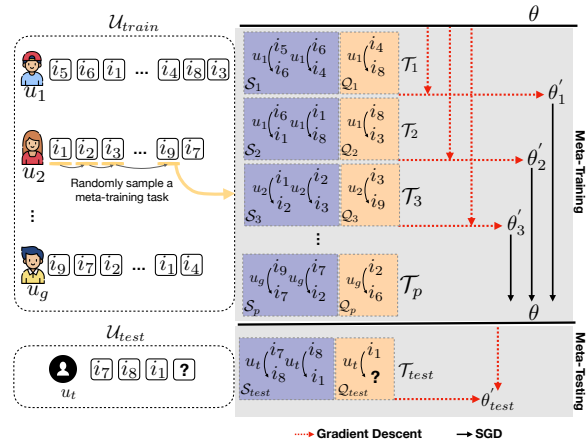}
\vspace{-0.1in}
\caption{Overview of the proposed \textit{MetaTL}.}
\vspace{-0.1in}
\label{fig:model}
\end{figure}

\subsection{Meta Transitional Learning}

\smallskip
\noindent\textbf{Few-shot Sequential Recommendation (RQ1).} Meta-learning aims to learn a model which can adapt to new tasks (i.e., new users) with a few training samples. To enable meta-learning in sequential recommendation for cold-start users, we formulate training a sequential recommender as solving a new few-shot learning problem (i.e., meta-testing task) by training on many sampled similar tasks (i.e., the meta-training tasks). Each task includes a $support$ set $\mathcal{S}$ and a $query$ set $\mathcal{Q}$, which can be regarded as the ``training'' set and ``testing'' set of the task. For example, while constructing a task $\mathcal{T}_n$, given user $u_j$ with initial interactions in sequence (e.g., $i_A\xrightarrow[]{u_j}i_B\xrightarrow[]{u_j}i_C$), we will have the a set of transition pairs $\{i_A\xrightarrow[]{u_j}i_B, i_B\xrightarrow[]{u_j}i_C\}$ as support and predict for the query $i_C\xrightarrow[]{u_j}?$. 

How can we generate a pool of meta-training tasks from data-rich users to mimic the targeted cold-start scenarios? Assume that we are focusing on predicting for cold-start users with $K$ initial interactions and want to predict for their $K+1^{th}$ interactions. To construct a meta-training task, firstly, we will randomly select a user $u_j$ from $\mathcal{U}_{train}$ and also randomly sample $K+1$ interactions from the user's logged interactions. With the $K+1$ interactions ordered chronologically ($i_1, i_2, ..., i_{K}, i_{K+1}$), we will have $i_1\xrightarrow[]{u_j}i_2, i_2\xrightarrow[]{u_j}i_3,..., i_{K-1}\xrightarrow[]{u_j}i_{K}$ in the support set and $i_{K}\xrightarrow[]{u_j}i_{K+1}$ for the query. 


\smallskip
\noindent\textbf{Meta transitional Learner (RQ2).}
Our goal is to gain the capability of learning the transition patterns to help predict the next interactions for a new user with only a few initial interactions.  

Firstly, for a task $\mathcal{T}_n$ sampled from user $u_j$, we want to retrieve the transitional dynamics with the transition pairs in support set $\mathcal{S}_n$. Let $i_{h}\xrightarrow[]{u_j}i_{t}$ denote a transition pair and $\textbf{i}_h$ and $\textbf{i}_n$ as the item embedding. Following \cite{chen2019meta}, then we can use a simple MLP network to represent the transitional information with item $i_{h}$ and item $i_{t}$, 
\begin{equation}
    \begin{aligned}
        \textbf{t}_{h,t} = \sigma \big(\textbf{W}(\textbf{i}_h||\textbf{i}_t) + \textbf{b} \big)\\
    \end{aligned}
\end{equation}
in which $\textbf{W}$ and $\textbf{b}$ is the trainable transform matrix and bias vector correspondingly. $\sigma(\cdot)$ denotes the Sigmoid activation function. 

Since there are multiple transition pairs in a support set $\mathcal{S}_n$ for the current task $\mathcal{T}_n$, we need to aggregate the transitional information from all the pairs to generate the final representation $\textbf{tr}_n$. As a straightforward solution, we take the average of all the transitional information from each transition pair in the support set $\mathcal{S}_n$:
\begin{equation}
    \begin{aligned}
        \textbf{tr}_n = \frac{1}{|\mathcal{S}_n|}\sum_{(i_{h}\xrightarrow[]{u_j}i_{t}) \in \mathcal{S}_n}\textbf{t}_{h,t},
    \end{aligned}
\end{equation}
in which $|\mathcal{S}_n|$ denotes the size of $\mathcal{S}_n$. After generating $\textbf{tr}_n$, we need a scoring function to evaluate its effectiveness in characterizing the transition pattern of the user. 
Following the idea in translation-based recommendation \cite{he2017translation}, a user can be viewed as the translation (or transition) between two consecutive items in the interaction sequence. Thus if $\textbf{tr}_n$ is effective in characterizing the transition pattern of a transition pair in $\mathcal{S}_n$ (i.e., $i_{h}\xrightarrow[]{u_j}i_{t}$), the translation $\textbf{i}_h + \textbf{tr}_n$ should be close to $\textbf{i}_h$. That is to say, the score $s(i_{h}\xrightarrow[]{u_j}i_{t}) = \| \textbf{i}_h + \textbf{tr}_n - \textbf{i}_h\|^2$ will have a small value. Then we calculate the marginal loss based on all the transition pairs in support $\mathcal{S}_n$ as: 
\begin{equation}
    \begin{aligned}
        \mathcal{L}_{\mathcal{S}_n} = \sum_{(i_{h}\xrightarrow[]{u_j}i_{t}) \in \mathcal{S}_n} [ \gamma + s(i_{h}\xrightarrow[]{u_j}i_{t}) - s(i_{h}\xrightarrow[]{u_j}i_{t}')]_{+}
    \end{aligned}
    \label{eqn:loss_cal}
\end{equation}
in which $[\cdot]_+ \triangleq max(0,\cdot)$ and $\gamma$ denotes the margin. Here $i_{t}'$ is a negative items without interaction from $u_j$. 

\smallskip
\noindent\textbf{Optimization and Fast Adaptation (RQ3).}
Next, we explain the procedure to optimize the model so that it can learn the transition pattern for new users with just a few interactions. We denote the meta model as a parameterized function $f_{\bm\theta}$ with parameters $\bm\theta$. Since we hope that the model can obtain a small value in $\mathcal{L}_{\mathcal{S}_n}$, we update the model $f_{\bm\theta}$ by minimizing the $\mathcal{L}_{\mathcal{S}_n}$ with one gradient step:
\begin{equation}
    \bm\theta_n' = \bm\theta - \alpha\nabla_{\bm\theta}\mathcal{L}_{\mathcal{S}_n}(f_{\bm\theta}),
    \label{eqn:update_nheta}
\end{equation}
in which $\alpha$ is the task-learning rate. Note that we can extend this to perform multiple-step gradient updates based on Eq. (\ref{eqn:update_nheta}). 

After updating $\bm\theta$ to be $\bm\theta_n'$ with the support set $\mathcal{S}_n$, we can generate the updated $\textbf{tr}_n$ with $f_{\bm\theta_n'}$. We evaluate its performance on the query set $\mathcal{Q}_n$ with loss $\mathcal{L}_{\mathcal{Q}_n}$, which can be calculated following Eq. (\ref{eqn:loss_cal}) with transition pair in $\mathcal{Q}_n$. Our goal is to determine the $\bm\theta$ that can work as a good initialization for each task. That is, it can minimize the loss on query sets across multiple meta-training tasks, 
\begin{equation}
\bm\theta = \min_{\bm\theta} \sum_{\mathcal{T}_n \sim p(\mathcal{T})} \mathcal{L}_{\mathcal{Q}_n}(f_{\bm\theta_n'}),
    \label{eqn:meta_loss}
\end{equation}
in which $p(\mathcal{T})$ is the distribution of meta-training tasks and can be obtained by randomly sampling meta-training tasks. To solve this equation, we can perform optimization via stochastic gradient descent (SGD), such that:
\begin{equation}
    \bm\theta \leftarrow \bm\theta - \beta\nabla_{\bm\theta}\sum_{\mathcal{T}_n \sim p(\mathcal{T})} \mathcal{L}_{\mathcal{Q}_n}(f_{\bm\theta_n'}),
\end{equation}
where $\beta$ is the meta step size.

When \textbf{testing} on a new user $u_{test}$, we will firstly construct the support set $\mathcal{S}_{test}$ based on the user's initial interactions. With Eq. ( \ref{eqn:update_nheta}), the model $f_{\bm\theta}$ is fine-tuned with all the transition pairs in $\mathcal{S}_{test}$ and updated to $f_{\bm\theta_{test}'}$, which can be used to generate the updated $\textbf{tr}_{test}$. Given the test query $i_{o}\xrightarrow[]{u_{test}}?$, the preference score for item $i_p$ (as the next interaction) is calculated as $- \| \textbf{i}_o + \textbf{tr}_{test} - \textbf{i}_p\|^2$.

\begin{table}[t]
\centering
\scalebox{0.9}{
\begin{tabular}{c|cccc}
\hline
 & \# Users & \# Items & \begin{tabular}[c]{@{}c@{}}Avg. Length \\ of Sequences\end{tabular} & \begin{tabular}[c]{@{}c@{}} Splitting \\ Timestamp\end{tabular}\\ \hline
\textbf{\textit{Electronics}} & 22,685  & 20,712 & 15.26 & 1/1/2014\\ \hline
\textbf{\textit{Movie}} & 26,933  & 18,855 & 28.97 & 1/1/2014\\ \hline
\textbf{\textit{Book}} & 90,892  &  58,250 &  27.81 & 1/1/2017\\ \hline
\end{tabular}%
}
\caption{Dataset Statistics.}
\label{tab:data}
\vspace{-0.15in}
\end{table}

\section{Experiments}
\subsection{Experimental Setup}
\smallskip
\noindent\textbf{Datasets.} We adopt three public real-world datasets: \textbf{\textit{Electronics}} is adopted from the public Amazon review dataset \cite{mcauley2015image}, which includes reviews ranging from May 1996 to July 2014 on Amazon products belonging to the ``Electronics'' category. \textbf{\textit{Movie}} is similarly drawn from the ``Movie'' category of the same Amazon review dataset. For both, we treat a user review as an interaction. \textbf{\textit{Book}} is scraped from Goodreads, a book platform for users to tag, rate and review books. We treat all these interactions on items equally. 

For all of the datasets, we filter out items with fewer than 10 interactions. We split each dataset with a corresponding cutting timestamp $T$, such that we construct $\mathcal{U}_{train}$ with users who have interactions before $T$ and construct $\mathcal{U}_{test}$ with users who start their first interactions after $T$. Summary statistics of the datasets can be found in Table \ref{tab:data}. When evaluating few-shot sequential recommendation for a choice of $K$ (i.e., the number of initial interactions), we keep $K$ interactions as initialization for each user in $\mathcal{U}_{test}$ and predict for the user's next interactions.


\smallskip
\noindent\textbf{Baselines.} Aside from a standard matrix factorization method  (\textbf{BPR-MF}) \cite{rendle2009bpr}, we compare the proposed MetaTL model with two categories of widely used recommendation models for Top-N recommendation: (i) \textit{Sequential recommendation baselines} utilize different methods to capture the sequential patterns in the interaction sequences of users. \textbf{TransRec} \cite{he2017translation} embeds items into a ``transition space'' and learns a translation vector for each user. \textbf{GRU4Rec} \cite{hidasi2015session}, \textbf{TCN} \cite{yuan2019simple} and \textbf{SASRec} \cite{kang2018self} rely on Gated Recurrent Units, the simple convolutional generative network, and the self-attention layers to learn sequential user behaviors, respectively. \textbf{BERT4Rec} \cite{sun2019bert4rec} adopts the bi-directional transformer to extract the sequential patterns and it is the state-of-the-art for sequential recommendation; (ii) \textit{Cold-start baselines} include methods focusing on providing accurate recommendations for cold-start users with limited information. \textbf{MeLU} \cite{lee2019melu} learns a user preference estimator model based on Model-Agnostic Meta-Learning (MAML), which can be rapidly adapted for cold-start users. We modify MeLU as in ~\cite{wei2020fast} to fit for the case without auxiliary information. \textbf{MetaCF} \cite{wei2020fast} learns a collaborative filtering (CF) model which can quickly adapt to new users. We adopt the version on top of NGCF \cite{NGCF19} for better performance.

\smallskip
\noindent\textbf{Evaluation Metrics.} In the experiment, each user only has one ground-truth item for testing. With the predicted scores, as in \cite{kang2018self,wei2020fast}, we rank the ground-truth positive item with 100 randomly sampled negative items. Mean Reciprocal Rank ($MRR$) indicates the rankings of the ground-truth items. We also evaluate the Hit Rate ($Hit$) for the top-1 prediction. $Hit@1=1$ if the ground-truth item is ranked top-1, otherwise $HR@1=0$. Also note that $HR@1$ is equal to the recall or NDCG for top-1 prediction.

\subsection{Overall Performance}

\begin{table}
\scalebox{0.9}{
\begin{tabular}{r|cc|cc|cc}
\hline
\multirow{2}{*}{} & \multicolumn{2}{c|}{\textbf{Electronics}} & \multicolumn{2}{c|}{\textbf{Movie}} & \multicolumn{2}{c}{\textbf{Book}} \\ \cline{2-7} 
 & Hit@1 & MRR & Hit@1 & MRR & Hit@1 & MRR \\ \hline
\textbf{BPR-MF} & 0.066 & 0.123 & 0.025 & 0.083 & 0.043 & 0.098\\
\textbf{TransRec} & 0.183 & 0.296 & 0.208 & 0.321 & 0.335 & 0.454 \\
\textbf{GRU4Rec} & 0.185 & 0.301 & 0.189 & 0.309 & 0.330 & 0.466 \\
\textbf{TCN} & 0.182 & 0.303 & 0.186 & 0.314 & 0.349 & 0.489 \\
\textbf{SASRec} & 0.193 & 0.318 & 0.211 & 0.345 & 0.347 & 0.488 \\
\textbf{BERT4Rec} & 0.200 & 0.323 & 0.220 & 0.351 & 0.369 & 0.513 \\ \hline
\textbf{MeLU} & 0.172  & 0.265  &  0.168 & 0.289  & 0.318  & 0.423  \\
\textbf{MetaCF} & 0.210$\dagger$ & 0.330$\dagger$ & 0.234$\dagger$& 0.365$\dagger$ & 0.398$\dagger$ & 0.528$\dagger$ \\
\hline \hline \textbf{MetaTL} & \textbf{0.224}  & \textbf{0.352}   & \textbf{0.258}  & \textbf{0.380}  & \textbf{0.420}  & \textbf{0.555}  \\ \hline
\end{tabular}%
}
\caption{Comparison of Different Models under $K=3$. The improvement of MetaTL is statistically significant compared with the next-best model with $p<0.05$}
\label{tab:result}
\vspace{-0.15in}
\end{table}

We compare the performance of MetaTL and the baseline models under $K=3$ and report the results in Table \ref{tab:result}. The best performing method in each column is boldfaced, and the second best method is marked with $\dagger$. MetaTL achieves the best performance under different evaluation metrics in all of the datasets, illustrating the effectiveness of MetaTL in providing accurate sequential recommendation for cold-start users with limited interactions. 

Starting from the simplest collaborative filtering (BPR-MF), we find that it performs weakly since it ignores the dynamic patterns in the user interaction sequences and fails to learn effective embeddings for cold-start users. TransRec also becomes ineffective in learning translation embeddings for cold-start users since there is insufficient data to update the embeddings. Then we compare the performance of various neural models for sequential recommendation. We can see that GRU4Rec and TCN perform the worst. SASRec and BERT4Rec (utilizing transformers to extract the sequential patterns) work better since they are able to aggregate the items with attention scores and thus obtain more informative representations for users with limited interactions.

MeLU and MetaCF are both meta-learning based methods for providing cold-start recommendations. Since MeLU requires side information for both users and items, we treat their historic interactions as the side information as in~\cite{wei2020fast}. Alas, MeLU is unable to obtain satisfying results since it designed for scenarios with abundant auxiliary user/item information which is absent in this case. Meanwhile we find that MetaCF can achieve the second-best performance for sequential recommendation, illustrating the importance of fast adaption in cold-start scenario. It still falls behind the proposed MetaTL since it is unable to learn the transition patterns for cold-start sequential recommendation. 


\subsection{Ablation Analysis}
\begin{figure}
\vspace{-0.1in}
    \centering
    \subfigure[\textbf{Movie}] 
    {
    \includegraphics[width=0.47\columnwidth]{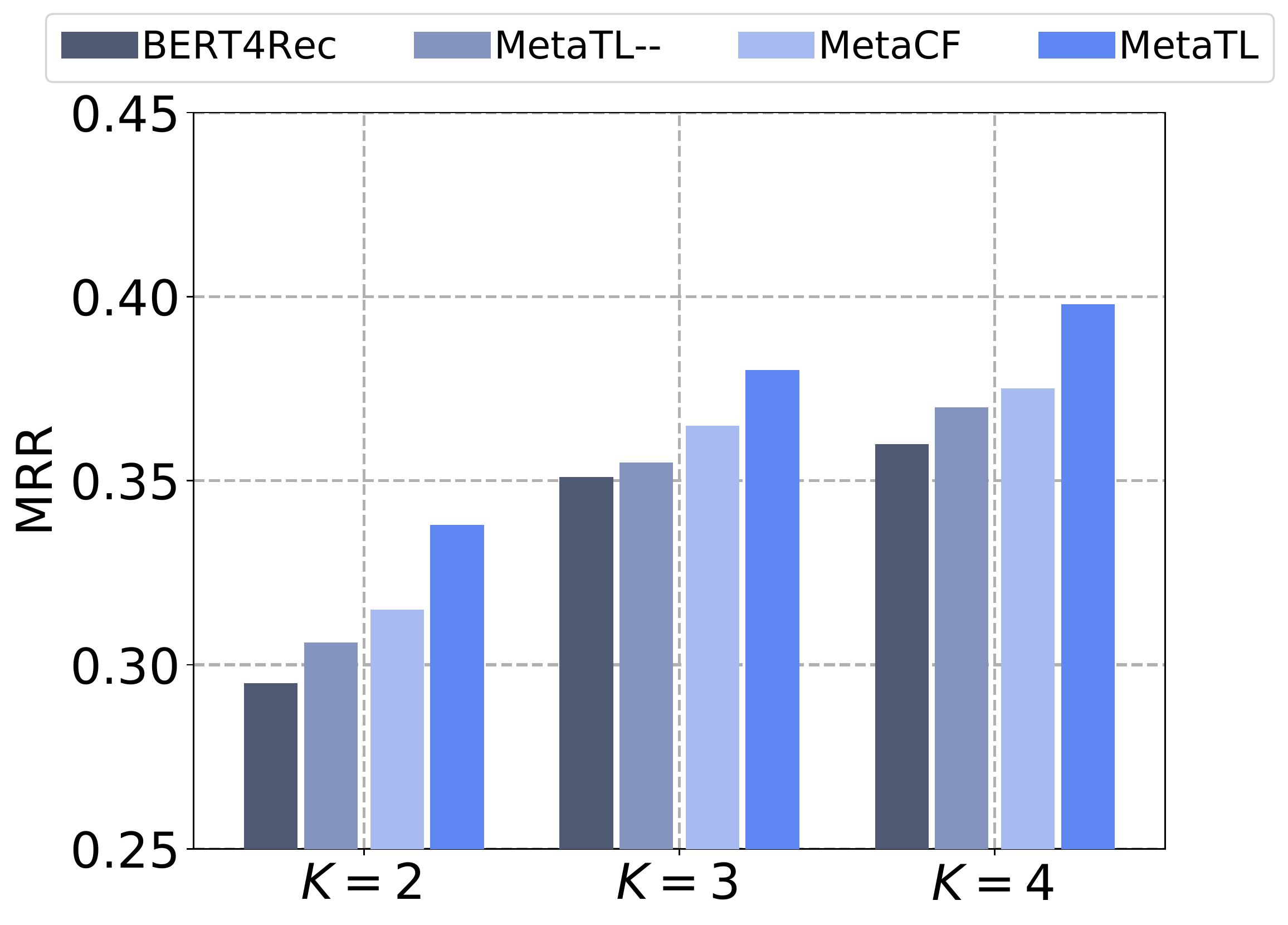}
    }
    \subfigure[\textbf{Book}]
    {
    \includegraphics[width=0.47\columnwidth]{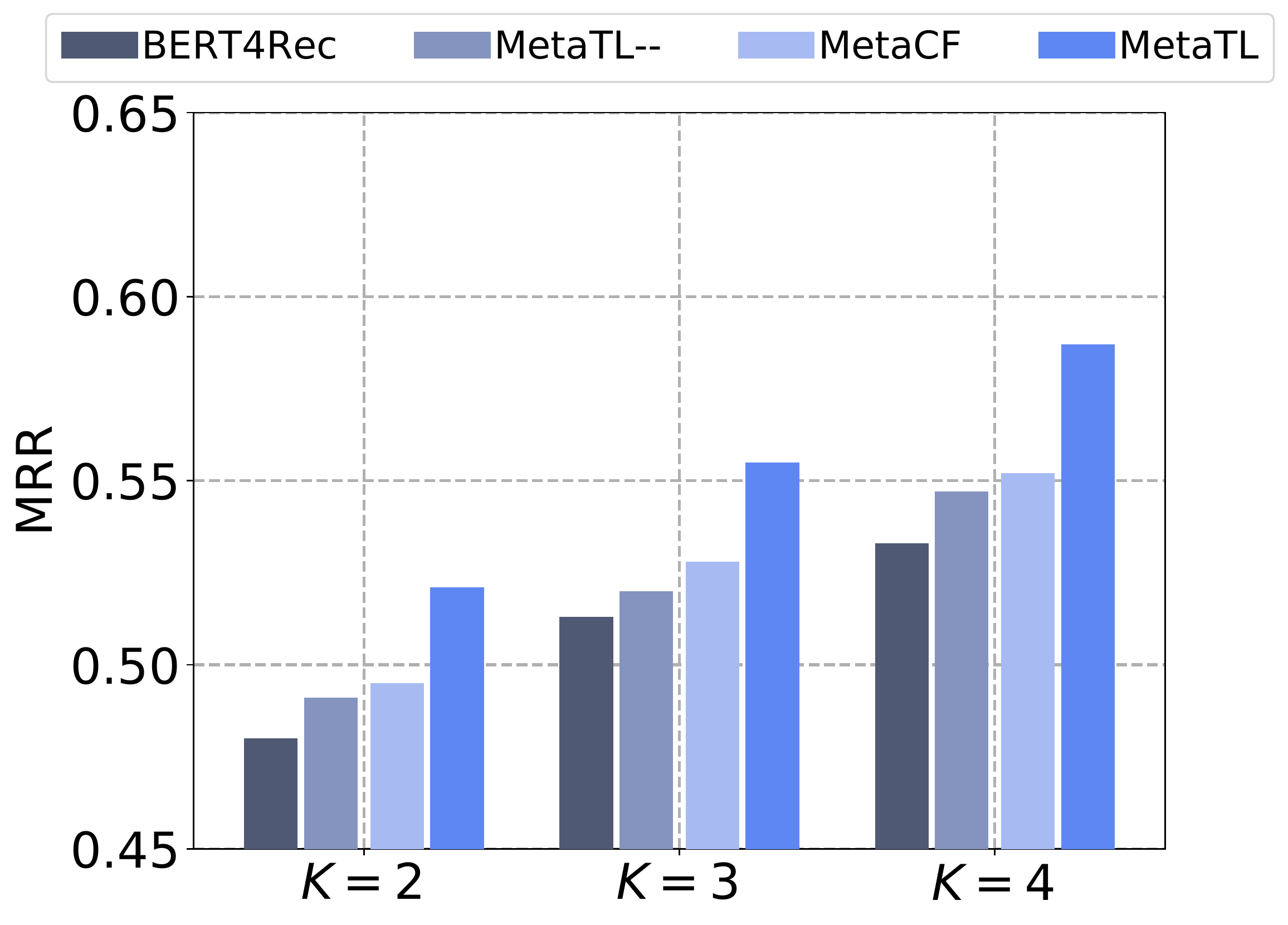}
    }
    \vspace{-0.05in}
    \caption{Comparison for different model variants \textit{w.r.t.} K.}
    \label{fig:ksz}
    \vspace{-0.05in}
\end{figure}

To further evaluate the effectiveness of the proposed MetaTL model, we compare it with its variants and some of the baselines under different K values (i.e., the number of initial interactions) in Figure \ref{fig:ksz}. Note that \textit{MetaTL}-- is the simplified version of MetaTL by removing the MAML optimization step. BERT4Rec is the state-of-the-art sequential recommendation method and MetaCF is the strongest cold-start baseline from our original experiments (and illustrates the performance of CF with meta-learning). 

In both datasets, BERT4Rec loses the prediction power on sequences consisting of only a few items and thus performs weakly in the cold-start sequential recommendation task. Even without the fast adaptation learning module, MetaTL-- can outperform BERT4Rec since it is carefully designed to learn the transition patterns on extremely short sequences. However, it still falls behind MetaCF, which demonstrates the necessity of training a model with fast adaptation in cold-start scenarios. With the well-designed optimization steps and meta transitional learner, the proposed MetaTL can further improve MetaTL-- and outperform both the state-of-the-art methods in sequential recommendation and cold-start user recommendation with different numbers of initial interactions. 

\section{Conclusion}
We propose a novel framework MetaTL to improve sequential recommendation for cold-start users. To enable the fast adaptation to cold-start users, we reformulate our task as a few-shot learning problem and adopt meta-learning for solving the problem. Powered by a translation-based architecture, the model is able to capture the transition patterns from the transition pairs. Meanwhile, given a pool of few-shot user preference transition tasks that mimic the targeted cold-start scenarios, MetaTL learns a model which can be adapted to new users with just a few interactions in a meta-learning fashion.  
With experiments on three real-world datasets, the proposed MetaTL can bring in significant improvement compared with the state-of-the-art methods.

\balance
\bibliographystyle{ACM-Reference-Format}
\bibliography{sample-bibliography} 

\end{document}